\documentclass[manuscript,screen,nonacm]{acmart}

\usepackage[acronym]{glossaries}
\usepackage{bussproofs}
\usepackage{subcaption}
\usepackage[capitalise]{cleveref}
\usepackage{graphicx}
\usepackage{url}
\usepackage{placeins}

\usepackage{array}
\usepackage{booktabs}
\usepackage{enumitem}
\usepackage{listings}
\newcolumntype{M}{>{\begin{math}}r<{\end{math}}}

\glsdisablehyper
\newacronym{abi}{ABI}{Application Binary Interface}
\newacronym{api}{API}{Application Programming Interface}
\newacronym{ast}{AST}{Abstract Syntax Tree}
\newacronym{cfd}{CFD}{Computational Fluid Dynamics}
\newacronym{ckd}{CKD}{Correlated K-Distributions}
\newacronym{dag}{DAG}{Directed Acyclic Graph}
\newacronym{di}{DI}{Debug Information}
\newacronym{dsl}{DSL}{Domain-Specific Language}
\newacronym{esn}{ESN}{Einsteinian Summation Notation}
\newacronym{ffi}{FFI}{Foreign Function Interface}
\newacronym{fpga}{FPGA}{Field-Programmable Gate Array}
\newacronym{gcm}{GCM}{Global Climate Model}
\newacronym{hls}{HLS}{High-Level Synthesis}
\newacronym{ir}{IR}{Intermediate Representation}
\newacronym{kpn}{KPN}{Khan Process Network}
\newacronym{lsp}{LSP}{Language Server Protocol}
\newacronym{mlir}{MLIR}{Multi-Level Intermediate Representation}
\newacronym{ods}{ODS}{Operation Definition Specification}
\newacronym{sem}{SEM}{Spectral Element Method}
\newacronym{ssa}{SSA}{Static Single-Assignment}
\newacronym{tpo}{TPO}{Tensor Product Operator}
\newacronym{wrf}{WRF}{Weather Research \& Forecasting Model}

\lstdefinestyle{codeblock}{
    basicstyle=\ttfamily\scriptsize,
    breaklines=true,
    columns=fullflexible,
    keepspaces=true,
    showstringspaces=false,
    upquote=true
}
\lstdefinestyle{cppblock}{
    style=codeblock,
    language=C++
}
\lstdefinestyle{pyblock}{
    style=codeblock,
    language=Python,
    numbers=left,
    xleftmargin=1cm
}
\lstdefinelanguage{mlir}{
    morecomment=[l]{//},
    morestring=[b]",
    morekeywords={assoc,reduce,using,yield,read,subscript,add,mul,get_static,container},
    sensitive=true
}
\lstnewenvironment{bigmlircode}{\lstset{basicstyle=\ttfamily\footnotesize,breaklines=true,columns=fullflexible,keepspaces=true,showstringspaces=false,upquote=true,language=mlir,mathescape=true}}{}
\newfloat{listing}{tbp}{lol}
\floatname{listing}{Listing}

\newcommand{\NumPy}{\texttt{NumPy}}
\newcommand{\HiSPEET}{HiSPEET}

\newcommand{\theLang}{EKL}
\newcommand{\theCompiler}{\texttt{eklc}}
\newcommand{\theDialect}{\texttt{ekl}}
\newacronym{theLang}{EKL}{Everest Kernel Language}

\begin{document}

\title[MLIR-native DSL compilers]{Demonstrating a Future for MLIR-native DSL Compilers on a NumPy-like Example}

\author{Karl F. A. Friebel}
\orcid{0000-0001-9534-3978}
\affiliation{%
    \institution{TU Dresden}%
    \city{Dresden}
    \country{Germany}
}
\email{karl.friebel@tu-dresden.de}

\author{Jascha A. Ohlmann}
\orcid{0009-0004-6751-3158}
\affiliation{%
    \institution{TU Dresden}%
    \city{Dresden}
    \country{Germany}
}
\email{jascha.ohlmann@tu-dresden.de}

\author{Jeronimo Castrillon}
\orcid{0000-0002-5007-445X}
\affiliation{%
    \institution{TU Dresden}%
    \city{Dresden}
    \country{Germany}
}
\email{jeronimo.castrillon@tu-dresden.de}

\begin{abstract}
    Compilers for general-purpose languages have been shown to be at a disadvantage when it comes to specialized application domains as opposed to their \gls{dsl} counterparts.
    However, the field of \gls{dsl} compilers features little consolidation in terms of compiler frameworks and adjacent software ecosystems.
    As a result, considerable work is duplicated, lost to maintenance issues, or remains undiscovered, and most \glspl{dsl} are never considered ``production-ready''.
    One notable development is the introduction of the \gls{mlir}, which promises a similar impact on \gls{dsl} compilers as LLVM had on general-purpose tooling.

    In this work, we present a \NumPy{}-like \gls{dsl} made for offloading numeric tensor kernels that is entirely \gls{mlir}-native.
    In a first for open-source, it implements all frontend actions and semantic analyses directly within \gls{mlir}.
    Most notably, this is made possible by our new dialect-agnostic \gls{mlir} type checker, created for the future of \glspl{dsl} in \gls{mlir}.
    We implement a simple, yet effective, parallel-first lowering scheme that connects our language to another \gls{mlir} dataflow dialect for seamless offloading.
    We show that our approach performs well in real-world use cases from the domain of weather modeling and \gls{cfd} in Fortran.

\end{abstract}

\maketitle

\section{Introduction}

A \gls{dsl} provides a concise and idiomatic representation for a specific class of problems.
By directly borrowing domain terminology, users are freed from obtuse coding patterns that are barriers to entry and readability.
With their limited scope, \gls{dsl} compilers can sacrifice generality for simplicity and still meet the same quality of result.
Unsurprisingly, there is a sprawl of \glspl{dsl} in the published research landscape.
However, few have been adopted widely enough to leave academic circles.

A proposed solution is to build \gls{dsl} compilers in \gls{mlir}, an extensible \gls{ir} framework.
This allows them to interoperate, letting \gls{dsl} authors benefit directly from the work of their peers.
\Gls{mlir} also reduces the dependency surface of the \gls{dsl} compiler, letting the wider \gls{mlir} user base help maintain it.
Although advertised as ``batteries included'', using \gls{mlir} properly (i.e., reusably) still requires lots of experience.

This paper presents two contributions to the field of \gls{dsl} compilers:
\begin{itemize}
    \item A blueprint for \gls{mlir}-native \glspl{dsl} (\cref{sec:ekl-front-end}), i.e., deeply embedding \glspl{dsl} in \gls{mlir} to reduce bespoke infrastructure and increase reusability.
    \item A minimal and safe extension of \gls{mlir} (\cref{sec:mlir-type-checking}) to introduce subtyping and a reusable deductive type checker, used by our \gls{dsl} frontend.
\end{itemize}
As a demonstration, we designed and implemented a \NumPy{}-like \gls{dsl} in \gls{mlir}, up to and including \gls{ast} level and diagnostics.
We implemented a parallel-first, pipeline-later lowering (\cref{sec:ekl-middle-end}) to dataflow circuits in MLIR~\cite{2024_Etna,2024_dfg_mlir} for targeting heterogeneous execution platforms~\cite{2023_Olympus}.
We applied the resulting compiler to optical depth calculation in RRTMG~\cite{2005_RRTMG}, and 4 \glspl{tpo} from the \HiSPEET{}~\footnote{\url{https://gitlab.hrz.tu-chemnitz.de/hispeet/hispeet}} \gls{cfd} library (\cref{sec:experiments}).
Notably, our compiler can target \glspl{fpga} as well as OpenMP.

In the evaluation, we focus on 5 kernels featuring \glspl{tpo} and other idioms profitable for parallelization and offloading.
Our \gls{dsl} was designed to enable a novel offloading strategy that supports \glspl{fpga}, while providing expressiveness and static type- and bounds safety.
Compared to other tensor \glspl{dsl}, imitating \NumPy{} is more familiar with experts in our target domains of weather modeling and \gls{cfd}.
Our \gls{dsl} also requires deductive type checking, for which no reusable \gls{mlir} implementation existed prior to this work.

\section{Background and Related Work}

To motivate our \gls{theLang} \gls{dsl}, this section introduces the \gls{ckd} method and \glspl{tpo}.
Guided by these examples, we provide an overview of optimizing tensor compilers that might be applied to the problem.
Picking an \gls{mlir}-native approach, we close the section by giving an introduction into \gls{mlir}.

\subsection{RRTMG and \glsentryshort{ckd}}\label{sec:ckd}

\begin{figure}[h]
    \centering
    \includegraphics[width=\columnwidth]{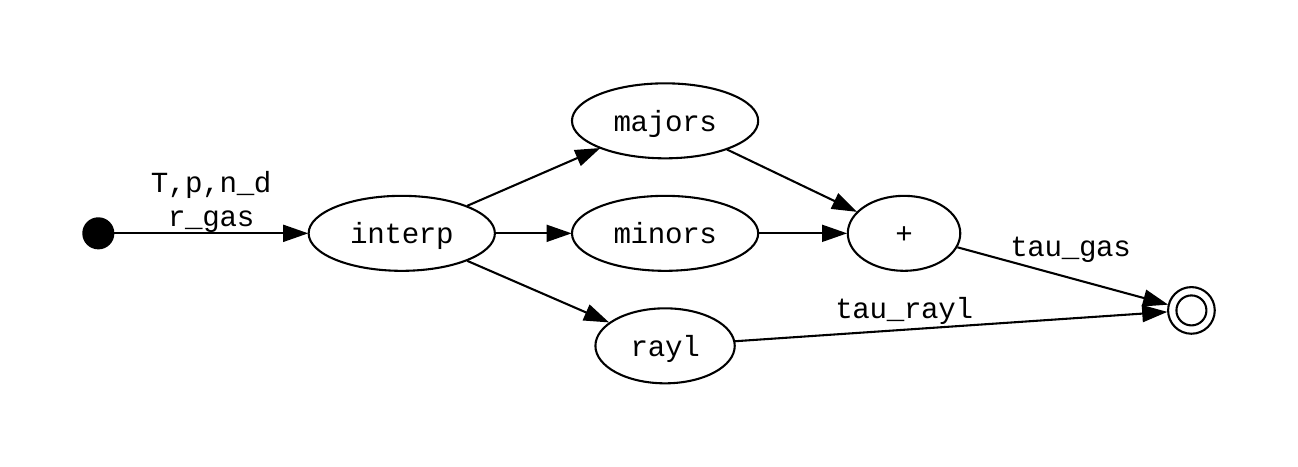}
    \vspace*{-1cm}
    \caption{High-level dataflow in the short-wave \glsentryshort{ckd} kernel.}\label{fig:ckd-df}
\end{figure}

Calculating the heating effects of global radiation is among the most time consuming tasks of \glspl{gcm}.
State-of-the-art radiative transfer models use the \gls{ckd} method to solve the underlying gas optics problem.
RRTMG~\cite{2005_RRTMG} and ecRAD~\cite{2016_ecRAD} are two important radiation schemes that feature the \gls{ckd} method (cf. \cref{fig:ckd-df}).

The RRTMG scheme developed by AER is a Fortran module of long pedigree, also used in \gls{wrf}~\cite{2019_WRF}.
Its hand-tuned implementation spans thousands of lines, and has inspired some GPU implementations~\cite{2016_CUDA_RRTMG,2021_GPUS_RRTMG}.
Notably, the original authors have created a new scheme~\cite{2019_RTE_RRTMGP} that ships their modernized \gls{ckd} implementation that uses OpenACC.

\begin{table}[b]
   \centering
   \begin{tabular}{>{\ttfamily}lccc}\toprule
        Kernel      & Fortran                   & \NumPy{}      & \theLang{}    \\\midrule
        taumol\_sw  & 3000 (500\(^\dagger\))    & 100           & 60            \\
        inv\_helm   & 80                        & ---           & 10            \\
        elliptic\_r & 100                       & ---           & 20            \\
        elliptic\_d & 200                       & ---           & 30            \\
        convection  & 600                       & ---           & 100           \\\bottomrule
   \end{tabular}\\
   \footnotesize{\(^\dagger\)In the modernized AER version.}\\[0.3\baselineskip]
   \caption{Lines of code for equivalent kernel snippets, up to a single significant digit, excluding comments and white space.}\label{tab:sloc}
\end{table}

\begin{listing}[h]
    \centering
\begin{lstlisting}[style=cppblock]
i_flav = spectra.bnd_to_flav[i_strato][:,i_bnd]
i_eta  = j_eta[i_lay,i_flav,...]
i_eta  = np.stack((i_eta,i_eta+1), -1)[:,None,:,:]
a = n_prime_mix[None,i_lay,i_flav,None,:,None]
b = f_major[None,i_lay,i_flav,:,:,:]
c = spectra.k_major[gptS:gptE,i_p,i_T,i_eta]
result[gptS:gptE,:] = np.sum(a * b * c, (2,3,4))
\end{lstlisting}
    \vspace*{-0.2cm}
    \caption{\texttt{taumol\_sw} major absorbers snippet in \NumPy{}.}\label{lst:tau-major-np}
\end{listing}

In the context of the EVEREST project~\cite{pilato_date21, pilato_date24}, we boiled down the original Fortran code by an order of magnitude (cf. \cref{tab:sloc}) by writing it as an idiomatic \NumPy{} program\footnote{\url{https://github.com/everest-h2020/wrf-plugin/tree/pyref/rrtmgp}}.
\Cref{lst:tau-major-np} shows a \NumPy{} snippet that computes the major absorber's contribution (the \texttt{taumol\_sw} kernel).
We use \NumPy{}'s ``Advanced Indexing'' feature for subscript-of-subscript expressions, e.g., indexing with \texttt{i\_eta}.
Line 7 performs linear interpolation by reducing over three axes that were stacked in line 3.
The result is reasonably fast \NumPy{} code, operating on array instead of scalar level.

\subsection{HiSPEET and \glsentryshort{tpo}s}\label{sec:hispeet}

\Gls{cfd} solvers interact strongly coupled quantities over large fields, leading to data dependencies that hinder parallelization.
By using Discontinuous Galerkin \glsentrylong{sem}s (DG-SEM)~\cite{2020_SDC}, these problems can be broken into localized computations~\cite{2005_SEM}.
\gls{tpo} formulations provide the basis for efficient implementations of such \glspl{sem}.
Instead of forming global matrices, operations are performed successively as one-dimensional transformations, reducing the computational complexity~\cite{2002_HOM}.

As an example, we selected HiSPEET\footnote{\url{https://gitlab.hrz.tu-chemnitz.de/hispeet/hispeet}}, a Fortran DG-\gls{sem} library for incompressible Navier-Stokes problems developed at TU Dresden.
We obtained the \gls{tpo} kernels labeled \texttt{inv\_helm}, \texttt{elliptic\_r}, \texttt{elliptic\_d} and \texttt{convection} from it.
For instance, the \gls{tpo} formulation of the volume integral of the convective flux from the Navier-Stokes \texttt{convection} operator is given by:

\begin{align}
    \vec{I}^{e,vol}_{ijk,c} &\approx
    (\mathcal{I}^{v,q}_{n,k} \otimes \mathcal{I}^{v,q}_{m,j} \otimes  D^{v,q}_{l,i}) M^{q,e}_{lmn}\vec{P}^{q,e}_{\xi  ,lmn} \nonumber\\ &+
    (\mathcal{I}^{v,q}_{n,k} \otimes D^{v,q}_{m,j} \otimes  \mathcal{I}^{v,q}_{l,i}) M^{q,e}_{lmn}\vec{P}^{q,e}_{\eta ,lmn} \nonumber\\ &+
    (D^{v,q}_{n,k} \otimes \mathcal{I}^{v,q}_{m,j} \otimes  \mathcal{I}^{v,q}_{l,i}) M^{q,e}_{lmn}\vec{P}^{q,e}_{\zeta,lmn} \label{eq:ins-conv-3d}
\end{align}
\subsection{Optimizing Tensor Compilers}\label{sec:tensor-compilers}

Tensor compilers target programs specified as expressions on tensors (rather: arrays / indexed families).
They are closely related to the fields of vectorizing- and polyhedral compilers, typically targeting linear algebra applications.

One family of tensor compilers focuses on operator implementation, i.e., the code generation of instructions as part of a subexpression.
TACO~\cite{2017_TACO} is a C++-embedded compiler that is known for its ability to exploit sparsity, which many practical applications have.
It provides a flexible and more convenient alternative to kernel libraries like MKL\footnote{\url{https://www.intel.com/content/www/us/en/developer/tools/oneapi/onemkl.html}}.

CFDlang~\cite{2018_CFDlang, rink_array19} and similar compilers leverage a domain-specific representation to exploit algebraic identities on tensors, e.g., accessible via meta-programming~\cite{rink_gpce18}.
Like many others, it relies on well understood scalar-level optimizations by deferring code generation, e.g., to LLVM.
Teckyl~\cite{2020_Teckyl} is a language and compiler implemented using \gls{mlir} that targets polyhedral code generation.
In an \gls{mlir} analog to above, its power derives from downstream dialects, in this case \texttt{affine}.

This family does not address high-level dataflow optimization, required for heterogeneous hardware deployment.
Since operator internals depend on data movement, we require compilers that transform both sides in the same \gls{ir} (also see \texttt{libomptarget} a.k.a. \texttt{llvm/offload}).

There is another compiler family that focuses mostly on dataflow optimizations and the rewriting of operator graphs.
They commonly apply constrained schedule manipulations like tiling and defer to operator-implementing compilers.
TVM~\cite{2018_TVM} and \gls{mlir} deep learning compilers like Triton\footnote{\url{https://github.com/triton-lang/triton}} and IREE\footnote{\url{https://iree.dev}} are prominent examples.
Among vendor-specific compilers, recombining known-optimal operator implementations, a.k.a. microkernels, is a popular approach.
Mojo\footnote{\url{https://www.modular.com/mojo}} enhances this with in-library design-space exploration.
Though not a compiler by itself, Torch-MLIR~\cite{TorchMLIR} is among the most commercially relevant \gls{mlir} frontends.
It models and imports the popular Python-embedded \gls{dsl} Torch Script for a range of \gls{mlir}-based AI flows.
Coincidentally, it started out as the \texttt{npcomp} \NumPy{} compiler experiment.

SPIRAL~\cite{2018_SPIRAL} is a compiler straddling both families, using an algorithmic representation that is both schedule agnostic and suitable for microkernel matching.
Operators dissolve into an expression form that fully expands their inherent parallelism before their compute is broken apart into optimized chunks.
This process is informed by hardware limits (e.g., vector width), and user-defined patterns.
We follow this approach, which it is a great fit for \gls{mlir}, because it can be incrementally improved by reusing foreign back-ends.

\subsection{Compiling \glsentryshort{dsl}s with \glsentryshort{mlir}}
\label{sec:mlir}

\Gls{mlir}~\cite{2021_MLIR} is an \gls{ir} often presented as ideal for \glspl{dsl}.
It is extensible, meaning that users can add their own semantics via dialects.
Dialects are sets of operations (ops), attributes and types that may appear in \gls{mlir}'s fixed structure.
There is also a set of ``core dialects'', which are abstractions maintained together by the community.
One of the essential core dialects is the \texttt{llvm} dialect that maps to LLVM-IR.
It acts as a common back-end to the whole \gls{mlir} ecosystem.
Creating a \gls{dsl} in \gls{mlir} can be as straightforward as adding a new dialect and a transitive conversion (i.e., lowering) to \texttt{llvm}.

\Gls{mlir} is also a compiler framework, providing a pass infrastructure and diagnostics engine.
Creating a compiler using \gls{mlir} is similar to LLVM: the compiler populates an \gls{ir} module, runs some pass pipeline on it, and then consumes the result to emit some artifact.
However, \gls{mlir} does not feature complete end-to-end compilers upstream.
Along with other recurring frontend matters, e.g. type checking, a feature-rich \gls{dsl} frontend presently still requires a substantial codebase that is not native to \gls{mlir}.

Dialects are the fundamental unit of reuse in \gls{mlir} and separate concerns under some completeness requirement.
For complex languages like C++, frontends may require multiple dialects for elaboration tasks before reaching executable semantics and code generation.
The ClangIR~\cite{ClangIR} project is creating such an \gls{mlir} stack for the C family of languages, but does not encompass the full \gls{ast}\footnote{C++ template instantiation is explicitly out-of-scope for ClangIR.}.

Still, the idea of using an \gls{mlir} dialect to model an \gls{ast} is not new, and in fact part of public \gls{mlir} training material.
We were unable to locate an open-source project doing this, however.
This may be because upstream \gls{mlir} lacks appropriate abstractions, or because legacy compilers often keep their \gls{ast} when ported to \gls{mlir}.

\section{\glsentryshort{mlir}-native \glsentryshort{dsl} implementation}\label{sec:ekl-front-end}

We designed \gls{theLang} to combine the idiomatic style of \NumPy{} with the benefits of a statically-typed language.
Its syntax and semantics mirror \NumPy{}'s, with a few exceptions we made for the sake of clarity and safety.

\Cref{lst:tau-major-ekl} shows the implementation of Lines 4-7 of \cref{lst:tau-major-np}.
Though supported, this snippet replaces implicit with explicitly named indices, which is less error-prone.
Note that all types and bounds are implicit, as allowed by the \gls{esn}.

\begin{listing}[h]
        \begin{lstlisting}[style=codeblock]
let tau_maj[b, x, G] =+ (dT, deta, dp)
    C_K_MAJOR[b, j_T[x]+dT, j_eta[x, b, dT]+deta, j_p[x]+dp, G]
  * f_major[x, b, dT, deta, dp]
  * f_mix[x, b, dT];
        \end{lstlisting}
    \vspace*{-1em}
    \caption{Major absorbers snippet in \theLang{}.}\label{lst:tau-major-ekl}
\end{listing}

Similarly straightforward is the translation of \cref{eq:ins-conv-3d} given by \cref{lst:ins-conv-3d-ekl}.
This direct translation to \gls{theLang} results in less than one-fifth of the equivalent Fortran lines of code (cf. \cref{tab:sloc}).
With the caveat of the column-major / row-major difference between Fortran and the C ABI of \gls{theLang}, \gls{theLang} kernels become drop-in replacements, linked by HiSPEET's CMake build system.

\begin{listing}[h]
        \begin{lstlisting}[style=codeblock]
let F_c_vol[c,e,k,j,i] =+ (l,m,n)
    D_vq[i,l]*I_vq[j,m]*I_vq[k,n]*M_qe[e,n,m,l]*P_qe[c,_0,e,n,m,l]
  + I_vq[i,l]*D_vq[j,m]*I_vq[k,n]*M_qe[e,n,m,l]*P_qe[c,_1,e,n,m,l]
  + I_vq[i,l]*I_vq[j,m]*D_vq[k,n]*M_qe[e,n,m,l]*P_qe[c,_2,e,n,m,l];
        \end{lstlisting}
    \vspace*{-1em}
    \caption{Convective flux volume integral in \theLang{}.}\label{lst:ins-conv-3d-ekl}
\end{listing}

One main benefit of strong types in \gls{theLang} is its static bounds safety and first-class \gls{esn} support.
However, these features add complexity to the front-end, which we will now show how to address natively in \gls{mlir}.

\subsection{Lexing \& Parsing}\label{sec:ast-in-mlir}

A text-based \gls{dsl} requires a parser that creates an in-memory representation for the compiler.
Most compilers will use a parser to create an \gls{ast}.
The process of transitioning between text and \gls{mlir} is usually handled by an \gls{mlir} target.
In this context, a target is an LLVM idiom that defines an importer (parser) and exporter (printer).
The first key aspect of our approach is to model the \gls{ast} directly in \gls{mlir}.

\begin{listing}[h]
    \begin{bigmlircode}
%54 = assoc (%i, %j, %k) {
  %81 = assoc (%i_77, %j_78, %k_79) {
    %83 = get_static @C_K_MAJOR -> !ekl.ref<f32[14, 14, 9, 60, 16]>
    %84 = read %83: !ekl.ref<f32[14, 14, 9, 60, 16]>
    %85 = subscript %2 (%j)
    %86 = add %85, %i_77
    %87 = subscript %39 (%j, %i, %i_77)
    %88 = add %87, %j_78
    %89 = subscript %15 (%j]
    %90 = add %89, %k_79
    %91 = subscript %84 (%i, %86, %88, %90, %k)
    %92 = subscript %51 (%j, %i, %i_77, %j_78, %k_79)
    %93 = subscript %53 (%j, %i, %i_77)
    %94 = mul %92, %93
    %95 = mul %91, %94
    yield %95
  }
  %82 = reduce %81 using (%arg6, %arg7) {
    %83 = add %arg6, %arg7
    yield %83
  }
  yield %82
}
    \end{bigmlircode}
    \vspace*{-1em}
    \caption{\glsentryshort{ast} for the major absorbers snippet. \glsentryshort{di} and the \texttt{ExpressionType} are omitted by the \glsentryshort{mlir} pretty printer.}\label{lst:tau-major-ast}
\end{listing}

\begin{figure}[h]
    \centering
    \includegraphics[width=0.4\textwidth]{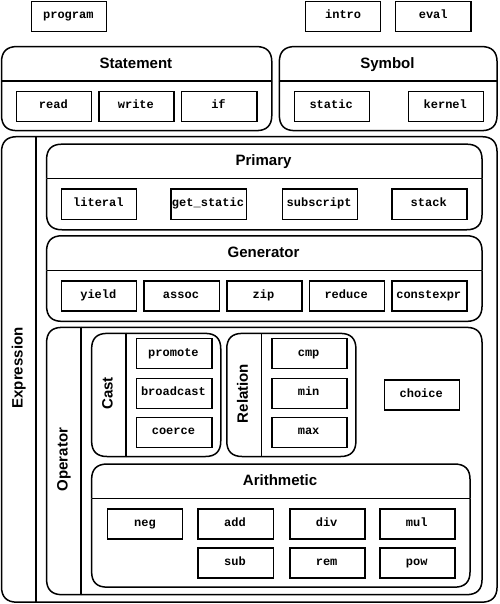}
    \vspace*{-1em}
    \caption{Operations in the \theDialect{} dialect.}\label{fig:ops}
\end{figure}

\Cref{lst:tau-major-ast} shows the \gls{mlir} that is parsed from \cref{lst:tau-major-ekl}.
In this representation, the operations are constructors of \gls{ast} nodes, with \gls{ssa} values connecting them.
This generalizes to any \gls{dsl}, as long as three requirements are met:
\begin{enumerate}[leftmargin=1em]
    \item All syntactically valid programs must have valid \gls{mlir} counterparts, i.e., ops in the dialect must support verification based on syntax only.
    Notice that \cref{lst:tau-major-ast} lacks concrete types, i.e., it is in syntactical form.

    \item A semantical \gls{ssa} form of the program must be constructible from the syntax, which should mutate the \gls{ir} as little as possible\footnote{This is also to preserve proper \gls{di} correspondence.}.
    To achieve this, every node that produces an expression must map \gls{ast} edges to \gls{ssa} values in \gls{mlir}, as opposed to, e.g., nested children.

    \item Since all \gls{ssa} values in \gls{mlir} require a type, it follows that every \gls{ssa}-referenceable \gls{ast} node must be assigned a ``syntactical'' type.
    The easiest way to do this is to introduce a single universal \texttt{ExpressionType}.
\end{enumerate}
\theLang{} is modeled by the \theDialect{} dialect and its ops shown in \cref{fig:ops}.
The \texttt{program} and \texttt{kernel} op are containers for nested ops and do not produce expressions.
\glspl{dsl} copying this approach can have straightforward parsers that simply append ops into the current container op whenever the corresponding syntax rule is applied.
We accomplished this using a trivial Bison parser, but projects like the \gls{mlir} \texttt{syntax} dialect are poised to eliminate the need for foreign parser generators in the future.

\theLang{} has contexts that must be evaluated at compile time, such as array bounds.
We use an ephemeral \texttt{constexpr} container to leverage the full power of an \gls{mlir}-native implementation.
Upon syntactical completion, this op holds an isolated \gls{theLang} program, which is immediately processed using our ordinary pipeline.
Thus, we can make use of transitive lowerings and foldings, including those of foreign \gls{mlir} dialects, to evaluate even more expressions at compile time.

\subsection{Type System}

\theLang{} models scalar and array types based on a subtype hierarchy.
Subtyping is a comparatively basic feature for a type system, but is not supported in status-quo \gls{mlir}.
We propose an extension to \gls{mlir} (cf. \cref{sec:mlir-type-checking}) that enables subtyping and type deduction to trivialize the implementation of \glspl{dsl} like \theLang{} and to improve \gls{ssa} composition in general.

\theDialect{} uses a fixed set of nominal types, modeled by \gls{mlir} types, and structural types, modeled by named requirements.
Named requirements, also called concepts, declare new type constraints that simplify operational semantics without increasing the number of type identities in the \gls{ir}.
The \theDialect{} dialect also uses a special nominal \texttt{IndexType} to enforce static bounds safety.

\begin{figure}
    \centering
    \includegraphics[height=0.4\textwidth]{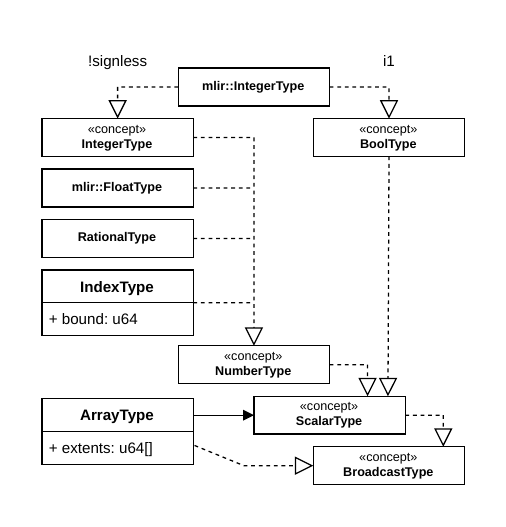}
    \vspace*{-1em}
    \caption{\glsentryshort{theLang}'s \texttt{BroadcastType} concept.}\label{fig:bcast-type}
\end{figure}

\Cref{fig:bcast-type} shows the \texttt{BroadcastType} concept, which is used in the definition of (implicitly) broadcasting ops.
In addition to this hierarchy, subtyping between two numeric types \(T\) and \(U\) is defined by their sets of values \(\texttt{vals}(\sigma)\)
\begin{equation*}
    T <: U \iff \texttt{vals}(T) \subseteq \texttt{vals}(U)
\end{equation*}
This relation can be decided based on the parameters of the two types.
Let \(\sqcup\) be a promotion operator that satisfies
\begin{equation*}
    V = T \sqcup U \implies V :> T \wedge V :> U
\end{equation*}
\theLang{} uniquely defines \(\sqcup_\mathbb{Q}\) over numeric types such that the smallest type is obtained.
This allows us to statically model \NumPy{} broadcast-and-promote semantics.

The \texttt{LiteralType}s are the set of types for literal expressions, and are annotated directly by the lexer.
Apart from scalars and strings, this set also contains the special identity (\texttt{:}), extent (\texttt{*}), ellipsis (\texttt{...}) and error (\texttt{?}) types.
\Gls{theLang} also defines \texttt{ReferenceType}s, which are used to establish its \gls{ffi}, i.e., \gls{theLang}'s C ABI.

\subsection{Type Checking}\label{sec:type-checking}

In a well-typed program, the types of all expressions are consistent under some set of typing rules.
Only well-typed programs are semantically valid, and the frontend must check that such an assignment exists, or reject the program.
In status quo \gls{mlir}, type checking happens during \gls{ir} verification, where each op typically checks its operand types.

Remember that \cref{lst:tau-major-ast} is in syntactical form, lacking concrete types, using the universal \texttt{ExpressionType} instead.
Our extension described in \cref{sec:mlir-type-checking} introduces typing rules to \gls{mlir}, allowing automated type deduction.
As a result of our \gls{ast} dialect design, this lets the dialect-agnostic type checking pass establish the semantic \gls{ssa} form.

Ops in the \theDialect{} dialect declare typing rules that mirror the semantics of \NumPy{}.
For example, using a partial \texttt{bcast} helper function, an addition expression in \NumPy{} is typed by the following rule

\begin{prooftree}
    \AxiomC{\(\Gamma \vdash l : T_l[s_l]\)}
    \AxiomC{\(\Gamma \vdash r : T_r[s_r]\)}
    \noLine
    \BinaryInfC{\(T = T_l \sqcup_\mathbb{Q} T_r \quad{} s = \texttt{bcast}(s_l, s_r)\)}
    \UnaryInfC{\(\Gamma \vdash l + r : T[s]\)}
\end{prooftree}

Here, \(T[s]\) names the broadcast type of scalar type \(T\) with shape vector \(s\).
We assume the premise is not fulfilled if \texttt{bcast} is undefined.
This rule is simply attached to the \texttt{add} op via an interface.
A well-typed expression under this rule hides implicit promotion and broadcasting.
This single-step type checking is complemented by an \gls{mlir} pattern that disentangles the cast, applied later in the middle-end.

\Cref{lst:tau-major-typed} shows the deduced types of index and array \gls{ssa} values from \cref{lst:tau-major-ast}.
In contrast to the rule shown above, the \gls{esn} rules make deductions about a definitions (the indices) from their uses (the subscripts).

\begin{listing}[h]
\begin{bigmlircode}
%54 = assoc (%i: !ekl<_13>, %j: !ekl<_59>, %k: !ekl<_15>) {
    %81 = assoc (%i_77: !ekl<_1>, %j_78: !ekl<_1>, %k_79: !ekl<_1>) {
      // ...
      yield %95: f32
    } -> !ekl.array<f32[2, 2, 2]>
    %82 = reduce %81: !ekl.array<f32[2, 2, 2]> using (%arg6: f32, %arg7: f32) {
      // ...
      yield %83: f32
    } -> f32
    yield %82: f32
} -> !ekl.array<f32[14, 60, 16]>
\end{bigmlircode}
    \vspace*{-1em}
    \caption{The major absorbers snippet after type inference.}\label{lst:tau-major-typed}
\end{listing}

\subsection{Benefits and Drawbacks}

\Glspl{dsl} that follow the same blueprint we used for \gls{theLang} benefit from fast and safe iterative design.
Little logic needs to go in the parser, and semantic analyses can be reused at the \gls{mlir} level.
Additionally, there is no code duplication when implementing the type system or syntactic sugar.
Our extensions to the infrastructure remove the hurdle on entry to \gls{mlir}.

On the other hand, \gls{mlir} is a complex framework that requires more boilerplate code than a bespoke \gls{ir}.
With an emphasis on integration and maintenance, however, we expect this to be amortized by reuse.
Complex languages may also require a hierarchy of frontend dialects (see \cite{ClangIR}), which may lead to issues with backtracing.

\section{Subtyping \& Type Checking in \glsentryshort{mlir}}\label{sec:mlir-type-checking}

This section describes our open-source\footnote{\url{https://github.com/KFAFSP/llvm-project/tree/mlir-type-system}} extension to the \gls{mlir} infrastructure to support type systems.
It introduces the concept of subtyping to \gls{mlir}, via a dialect interface.
This relation subsumes the value substitutability rule, minimally changing the core of \gls{mlir}.
To allow for efficient in-place \gls{ir} mutation, we extend substitutability to transmutability.
We implemented a dialect-agnostic type checking pass, which runs a fix-point algorithm on op typing rules advertised by a new op interface.

\subsection{Status-quo typing in the \glsentryshort{mlir} \glsentryshort{ssa} \glsentryshort{ir}}
Every value in \gls{mlir} requires a type.
Though \gls{mlir} has some built-in types, its core remains unopinionated.
Dialects can introduce new types, freely assign semantics to them, and even reinterpret foreign types in their semantics.
An \gls{mlir} type is simply a family of compile-time attributes with static lifetime and a notion of definitional equality.

There have been previous attempts at embedding richer type systems into \gls{mlir}, with features such as type abstractions.
For example, CIRCT module parameters are upstream type abstractions based on a simple but unsafe\footnote{Type variables are not scopable; definitionally equal types can be distinct.} workaround.
Safe approaches embed type expressions into \gls{mlir} by creating self-contained reasoning frameworks, which may bypass \gls{mlir} types completely.
This has been comprehensively demonstrated by Rise~\cite{2021_Rise} and MimIR~\cite{2025_MIMIR}.
However, these embeddings must remain closed, i.e., foreign expressions and types must be encapsulated.

All such embeddings lose the desired orthogonality of \gls{mlir} dialects, because they introduce composition hazards.
The alternative is to modify the \gls{mlir} core to add first-class support for rich type systems.
For features like dependent types, this requires sweeping changes to the \glsentryshort{abi}, \glsentryshort{api}, runtime and memory consumption of \gls{mlir} that its users care about and maintainers defend.
The majority of \gls{mlir} users does not desire for it to become a theorem prover itself.
Such a change would run a high risk of fracturing the project.

Our extension is informed by the de-facto use of \gls{mlir} and its development philosophy.
The implementation described in this section concedes that any enrichment of types in \gls{mlir} is limited by compatibility issues and non-functional constraints on \gls{mlir}.
Instead, we propose a reasonable patch that minimally modifies the spec and existing code of \gls{mlir} in a way that preserves the status quo while providing useful reasoning power and allowing gradual adoption.

\subsection{The \texttt{DialectTypeSystemInterface}}

Values in \gls{mlir} are considered substitutable when their types are definitionally equal.
Although this is an implicit contract that can be violated by op verifiers, it is one of the few invariants that is opinionated in the core of \gls{mlir}.
Subtyping relaxes this constraint throughout the core, touching on three core interfaces.

To introduce an extensible subtype relation to \gls{mlir} core, a new component is added to it such that:
\begin{itemize}
    \item It decides value substitutability.
    \item It is extensible via dialect loading.
    \item The load order of independent dialects can not matter.
    \item Dialects can disagree on type semantics.
    \item It does not change the behavior of existing compilers.
\end{itemize}

This is realized using the \texttt{DialectTypeSystemInterface} that is attached to dialects at load time.
Attaching to dialects rather than ops reduces the performance impact, assuming that all ops in a dialect agree on the same subtype relation.
This interface is an opt-in mechanism that allows querying the subtype relation and performing promotions.
It is also a starting point for future extensions.

The status quo is preserved in the \texttt{EqualityTypeSystem} default implementation.
All implementations must uphold reflexivity and transitivity, so that performance degradation is avoided in compilers that do not use this new feature.
Modifications to the core interfaces ensure that opted-in dialects get the expected behavior when dealing with control flow edges in their own \gls{ir}.

For performance reasons, we extend substitutability with transmutability.
During deduction, a common scenario is that a more concrete type is found for \(v\) that satisfies all of its current users.
Ideally, the type of the value \(v\) is updated in-place, avoiding costly \gls{ir} mutations.
Using our interface, we can now determine when this transmutation is legal.

Transmutability is more involved than evaluating the subtype relation, because of \texttt{BlockArgument}s.
These are values defined not by operations, but blocks contained within them.
In practice, we follow the control flow interfaces in addition to \gls{ssa} dataflow.
Since operations may have attributes synced to block arguments, e.g., function signatures, we allow customizing transmutation.
As before, transmutation requires opt-in, to preserve backward compatibility.

\subsection{The \texttt{TypeCheckOpInterface}}

\Gls{mlir} core features an \texttt{InferTypeOpInterface}, which is the existing type deduction mechanism.
It is specially privileged by \gls{mlir}, as it is known to the \gls{ods} and its parsers.
Primarily, it can be used to infer the result types of a yet unconstructed op from constructor arguments.
This allows omitting the types of op results in \gls{mlir}'s textual format.
It introduces a per op verifier for status quo type checking.

This interface is unsuitable for us, because it does not capture the concept of typing rules adequately.
And, while a typing rule subsumes it, its privileged status makes it somewhat orthogonal.
It serves a specific purpose in the context of \gls{ods} and its automatically generated C++ template builders.
We conclude that ops with syntactically inferable types should continue to use this interface.

We introduce the \texttt{TypeCheckOpInterface} to advertise typing rules.
Ops must override the \texttt{typeCheck} method, which evaluates the typing rule.
Implementations can also override a set of methods that control transmutation of results and block arguments.
The interface also adds a per op verifier to work with status quo type checking.
This feature is preserved by checking the rule locally against the current state of the \gls{ir}, avoiding further code duplication.

The \texttt{typeCheck} method takes an \texttt{AbstractTypeChecker} and returns an optional \texttt{Contradiction}.
The type checker is used to observe the current context and make deductions, i.e., add new constraints.
The contradiction result is used to implement rich diagnostics that explain when type checking gets stuck.
The typing rule can be invoked multiple times during checking, whenever the context of the operation changes.

\Cref{lst:type-check-add} shows the typing rule implementation shared by the arithmetic operations of the \theDialect{} dialect.
We use an adaptor to extend \texttt{AbstractTypeChecker} with dialect-specific functions, like \texttt{broadcastAndPromote}.

\begin{listing}[h]
    \begin{lstlisting}[style=cppblock,basicstyle=\ttfamily\footnotesize]
TypeCheckingAdaptor adaptor(typeChecker, *this);

ArithmeticType promotedTy;
if (auto contra = adaptor.broadcastAndPromote(
    adaptor.getOperands(),
    promotedTy,
    "arithmetic type"))
  return contra;

if (llvm::isa<ekl::IndexType>(promotedTy.getScalarType())) {
  // Arithmetic operations need to properly update the upper
  // bounds on the types of index values they produce...
}

return adaptor.meet(adaptor.getResult(0), promotedTy);
    \end{lstlisting}
    \caption{\texttt{typeCheck} code for arithmetic \theDialect{} ops.}\label{lst:type-check-add}
\end{listing}

\subsection{The \texttt{FixPointTypeChecker}}

\begin{figure}
    \centering
    \includegraphics[width=\columnwidth]{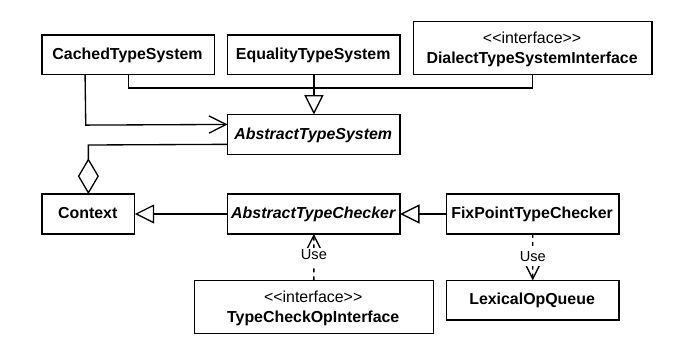}
    \caption{MLIR Type System extension overview.}\label{fig:mlir-type-system}
\end{figure}

\Cref{fig:mlir-type-system} shows an overview of the classes that were added to the core of \gls{mlir}.
The \texttt{AbstractTypeChecker} provides access to the typing context and handles conflicts that arise during rule application.
With the \texttt{FixPointTypeChecker}, we provide an implementation which applies typing rules until a fix-point or a contradiction is reached.
Future implementations may override this behavior while still consuming the same typing rules.

The core mechanism for deductions are type \texttt{Bound}s.
The context allows querying the bounds of values in the current \gls{ir} or type checking state.
A bound names a set of types using a sample and a subtype relation predicate on it.
The predicates are ordered from most to least restrictive (unattainable \(>\) equivalent \(>\) lower \(>\) upper).
The type checker meets bounds on values to determine the strictest, which is a commutative operation.
Thus, the subsumption rule is part of the infrastructure itself.

Dialect-agnostic type checking is implemented as an \gls{ir} pass that drives the fix-point type checker.
When the fix-point is reached, the pass atomically applies all deductions to the \gls{ir}.
By sampling the most restrictive bound, a type is found for each value, which is materialized in the \gls{ir} via transmutation and/or casting, depending on the users.
Using our infrastructure, the \gls{ir} is able to remain in a verified state throughout, which is important when it is used as an \gls{ast}.

The fix-point algorithm keeps a priority queue of rules that need to be (re-)applied.
Since these rules are given by ops, the queue is ordered in canonical lexical order of the \gls{ir}, which also helps reduce the number of iterations for deductions along \gls{ssa} edges.
Each iteration, a rule is dequeued and applied.
Assuming no contradictions are found, the new deductions are recorded into the state.
Deductions that refine the type of a value cause its users to be invalidated.

Because these \gls{ssa} values form a dataflow \gls{dag}, this process will terminate.
This termination guarantee is also preserved when rules explicitly invalidate child ops, as \gls{mlir} is also a \gls{dag}.
However, to accommodate idiomatic structural \gls{mlir} patterns, some parent ops also need to be invalidated.
Consider the following idiomatic \gls{mlir} example, where \texttt{\%i} is a capture operand and \texttt{\%j} its receiving block argument.
\begin{bigmlircode}
%r = container (%j = %i) {
    yield %j
}
\end{bigmlircode}
During inference, \texttt{container} will deduce the type of \texttt{\%j} to be the type of \texttt{\%i}.
This causes \texttt{yield}'s rule to be applied, but that op does not own the result \texttt{\%r}.
As a result, \texttt{yield} must propagate the invalidation to its parent, which can constrain \texttt{\%r}.
In this example, no infinite loop occurs, because the types of \texttt{\%j} and \texttt{\%r} are independent.
We hope that, in future, safe typing rule code is generated declaratively by \gls{ods}.

\section{The \theLang{} Middle-end}\label{sec:ekl-middle-end}

In \cref{sec:tensor-compilers}, we introduced the parallel-first, pipeline-later strategy for compiling tensor programs.
This strategy assumes that the optimal solution can be found from a maximum-parallel normalization form by gradually removing parallelism guided by the targets resources, e.g., hardware vector size.
Lowering \gls{theLang} programs to executable target code follows this strategy, preserving target independence for as long as possible.

\subsection{Program Normalization}

The first step is to normalize the program into a map-reduce form.
This is achieved using the inherently parallel generator expressions (cf. \cref{fig:ops}), a core design feature of \gls{theLang}.
The \texttt{assoc} generator produces an array by associating tuples of an index space with values returned by a child functor.
The \texttt{zip} generator produces an array by combining the values of its operands at matching index tuples using a child functor.
The \texttt{reduce} generator produces a scalar from an array by applying a reduction functor.

The first three stages of our pipeline establish this parallelized normal form.
Stage 1 is a cursory operator simplification, which rewrites compound expressions within \theDialect{}.
Stage 2 resolves the debt incurred by our single-step type checking (cf. \cref{sec:type-checking}), making implicit casts explicit.
Stage 3 rewrites array expressions, resulting in a mix of the embarassingly parallel \texttt{assoc} and \texttt{reduce} generators through aggressive inlining and code motion.

Examples for stage 2 include materializing broadcasts and type promotions.
Another example are \NumPy{} ellipses \texttt{...} subscripts, which expand to zero or more identity indexers \texttt{:} as needed.
Single-step type checking has left us with other such implicit expansions in \texttt{subscript} expressions, which are made explicit here.

\begin{figure}[ht!]
    \centering
    \begin{subfigure}{\columnwidth}
        \includegraphics[width=\columnwidth]{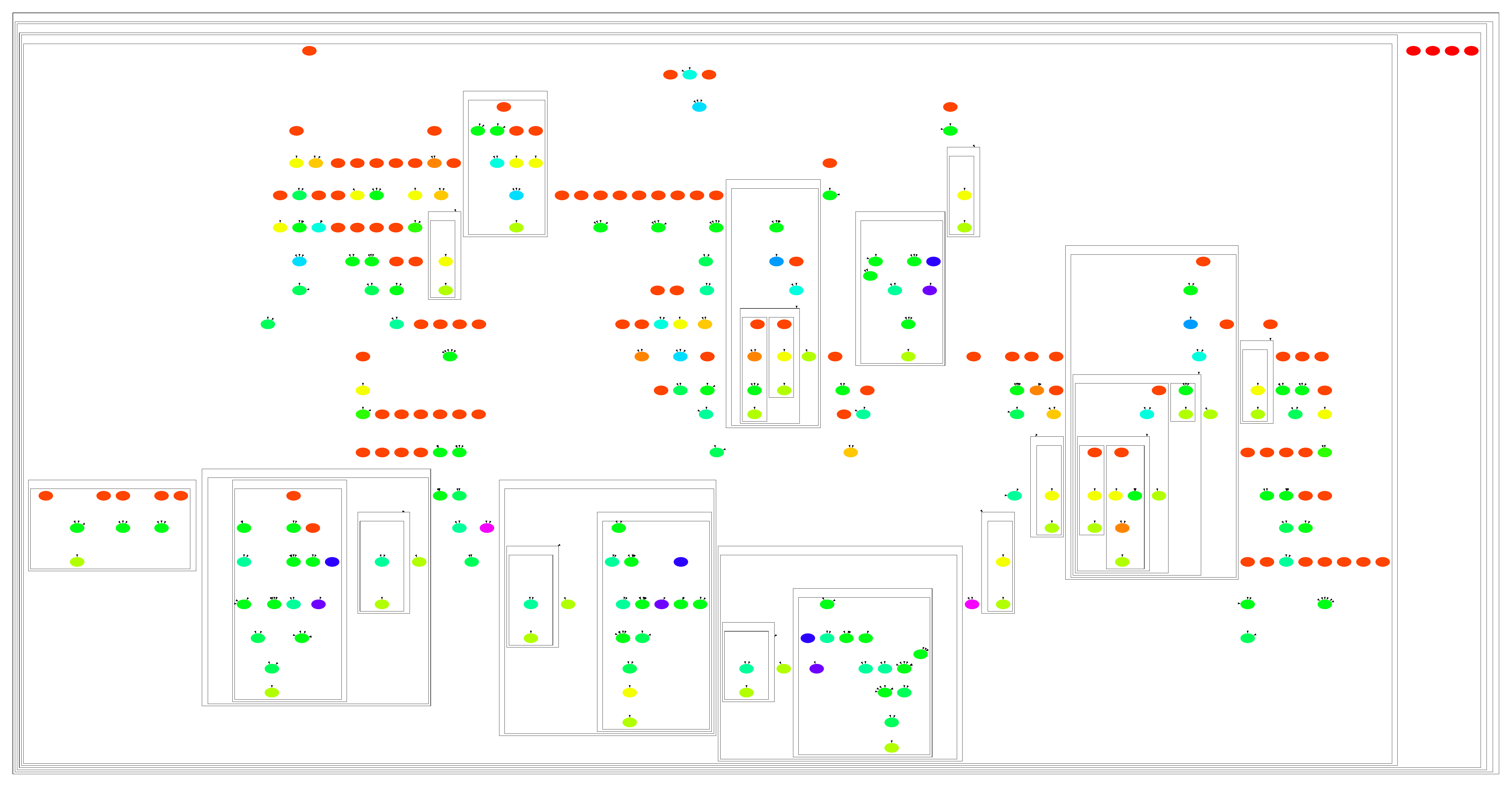}
        \caption{In \glsentryshort{ast} form.}
    \end{subfigure}
    \begin{subfigure}{\columnwidth}
        \includegraphics[width=\columnwidth]{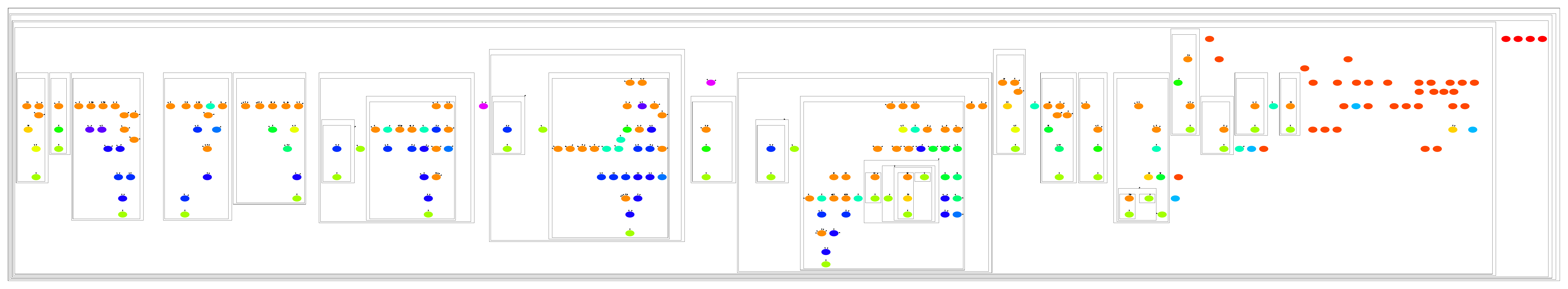}
        \caption{In the parallel normalization form.}\label{fig:parallel-minimap}
    \end{subfigure}
    \begin{subfigure}{\columnwidth}
        \includegraphics[width=\columnwidth]{figures/tensor.pdf}
        \caption{Implemented for a CPU target.}\label{fig:cpu-minimap}
    \end{subfigure}
    \vspace*{-1.5em}
    \caption{\glsentryshort{mlir} minimap of the \glsentryshort{ckd} kernel program.}\label{fig:middle-end}
\end{figure}

\Cref{fig:middle-end} illustrates the \gls{mlir} program structure after the relevant pipeline stages.
Dots represent ops in the program, arranged by their \gls{ssa} edges, nested into region boxes.
Since our ops are streamable, this provides a sketch of the dataflow in our program.
In the normalized form (cf. \cref{fig:parallel-minimap}), we can now observe a more fine-grained structure than \cref{fig:ckd-df}.
This is because the program was reduced to its true dependencies, i.e., the minimal constraints on the program schedule were exposed.

\subsection{Operator Simplification}

The program is then further simplified by (re-)introducing additional dependencies that trade off parallelism for time complexity and arithmetic intensity.
This includes code motion, such as hoisting and sinking of side-effect free operations.
An advantage of using \gls{mlir} to implement our \gls{dsl} is that many of these utilities are already available and reusable in our pipeline.
Custom additions include lowering our infinite-precision rational numbers to machine types, after exploiting some arithmetic properties.

One of these optimizations that is strictly beneficial is reduction lifting.
In \glspl{tpo}, reductions of tensor products can be rearranged using distributivity of operators to reduce the loop nesting.
In the following example, the naive time complexity can be reduced by a factor of \(p^2\):
\begin{align*}
    t_{ijk} &= \sum_{l=0}^{p} \sum_{m=0}^{p} \sum_{n=0}^{p} S^T_{li} \cdot S^T_{mj} \cdot S^T_{nk} \cdot u_{lmn} \\
            &= \sum_{l=0}^{p} S^T_{li} \sum_{m=0}^{p} S^T_{mj} \sum_{n=0}^{p} S^T_{nk} \cdot u_{lmn}
\end{align*}
These optimizations are usually unsafe for floating-point numbers, as their operations are neither associative nor distributive.
Our type system, however, is well-suited for this change.

We added \texttt{if} expressions and statements to \gls{theLang} as an escape hatch for ease of use and parity with the Python world.
Still, whenever possible, we eliminate or dissolve these ops based on known condition or result expressions.
Additionally, side-effect free \texttt{if} expressions are rewritten to parallelizable \texttt{choice} expressions.

\subsection{Operator Implementation}

We provide an operator implementation lowering via the upstream \gls{mlir} dialects \texttt{tensor}, \texttt{linalg}, \texttt{arith} and \texttt{affine}, which is enough to enable code generation via LLVM.
For more control over pipelining the operators, we use a pass that lifts all dataflow dependencies into the \texttt{dfg-mlir} dialect~\cite{2024_dfg_mlir}.
This can also be used to offload stages to heterogeneous compute devices, such as \glspl{fpga} via Bambu~\cite{2021_Bambu} and Olympus~\cite{2023_Olympus}.

While the pipeline is target-agnostic up until this point, it must specialize for the desired hardware target now.
To allow reuse of existing \gls{mlir} back-ends, computations must be exposed so that they can be mapped \(1:1\) onto target instructions.
Some program constructs are not yet efficiently mapped to CPU targets, as we initially only targeted \glspl{fpga}.

In particular, the \gls{fpga} target trades speed for area when parallel (re-)computations are combined with muxers, where the CPU prefers branches.
For the CPU, we currently apply a simple producer-consumer fusion to minimize reuse distance.
\Cref{fig:cpu-minimap} shows the result, where the major and minor absorber parts are clearly visible as two big nests.
The \texttt{rayl} part was absorbed into the minor absorbers, due to its low complexity and reuse distance.

Another advantage of using the \gls{mlir} ecosystem is our ability to target an OpenMP back-end with ease.
Introducing a single compiler flag, we are able to automatically extract parallel work share regions for the OpenMP runtime.
This back-end can also be used to implement and test the dataflow network for \gls{fpga} offloading locally.
However, the current state of \gls{mlir} bufferization means that the quality of result is still poor for CPU execution.
While dynamic memory allocations are rewritten to streams on the \gls{fpga}, the same is not applicable to the cached memory model of CPUs.

\section{Evaluation}\label{sec:experiments}

\newcommand{\us}[1]{\ensuremath{#1\,\mu{}s}}
\newcommand{\ms}[1]{\ensuremath{#1\,ms}}
\newcommand{\xs}[1]{\ensuremath{\times #1}}

\begin{table*}
   \centering
   \begin{tabular}{>{\ttfamily}lrrrcrc}\toprule
                    & \multicolumn{2}{c}{\texttt{gfortran}} & \multicolumn{4}{c}{\theCompiler{}}\\
                         \cmidrule(lr){2-3}                           \cmidrule(lr){4-7}
        Kernel      & seq.          & par.           & \multicolumn{2}{c}{seq.}  & \multicolumn{2}{c}{par.}  \\\midrule
        taumol\_sw  & \us{160}      & ---\(^\dagger\)& \us{140}      & \xs{1.1}  & \multicolumn{2}{c}{---\(^\dagger\)} \\
        inv\_helm   & \ms{8.73}     & \ms{13.0}      & \ms{5.36}     & \xs{1.6}  & \ms{40.3}     & \xs{0.3} \\
        elliptic\_r & \ms{44.5}     & \ms{12.4}      & \ms{46.4}     & \xs{0.9}  & \ms{173}      & \xs{0.07} \\
        elliptic\_d & \ms{110}      & \ms{35.5}      & \ms{252}      & \xs{0.4}  & \ms{71.7}     & \xs{0.5}  \\
        convection  & \ms{494}      & \ms{133}       & \ms{899}      & \xs{0.6}  & \ms{218}      & \xs{0.6}  \\\bottomrule
   \end{tabular}\\
   \footnotesize{\(^\dagger\)\gls{wrf}'s RRTMG processes each column single threaded.}\\[0.3\baselineskip]
   \caption{Latencies \& speedups for the kernels.}\label{tab:results}
\end{table*}

We conducted a series of benchmarks on a Ubuntu 22.04 server running on a Ryzen 9 7900X3D@3.8GHz.
As target for hardware synthesis, we selected the Xilinx Alveo U55C datacenter \gls{fpga} platform.
We validated the results of the kernels against the baseline, and measured their latencies.

\Cref{tab:results} gives the baseline under \texttt{gfortran} and ours under \theCompiler{}.
The ``seq.'' and ``par.'' columns show single-thread and 8-thread OpenMP execution respectively.
Time measurements are provided with 3 significant digits, and speed-ups with 2 significant digits.
Speed-ups less than \(\xs{1}\) indicate an increase in latency.

For specific applications, our naive approach turned out to be competitive with the expert-optimized code.
While further target specialization in \theCompiler{} is necessary to address the other kernels, we also identified key deficiencies in the \gls{mlir} back-ends reused by us.

\subsection{Experiment Setup}

Evaluation of the \texttt{taumol\_sw} kernel was performed using a full \gls{wrf} predicition run, the configuration and input data of which was provided by our partners.
In the commonly used build configuration \texttt{dmpar} of \gls{wrf} that we used, the kernel executes single-threaded.
Instead, the workload is spatially distributed across threads.
Our kernel is dynamically linked into \gls{wrf} as an open-source\footnote{\url{https://github.com/everest-h2020/wrf-plugin}} plugin, allowing the baseline to be swapped in and out.

To easily reproduce the experiment, also on \gls{fpga}, we derived controlled experiment conditions from these runs.
Since the kernel exhibits data-dependent timing, we recorded realistic replay data from the original simulation runs.
We then replay this data to the kernel using a test adapter that calls our plugin library.
For \texttt{taumol\_sw}, we averaged the latency of 1 million single-column runs, which are the smallest unit of work \gls{wrf} can distribute.

The HiSPEET kernels were tested using a Fortran application that directly integrates the \gls{dsl}.
The baseline HiSPEET implementation also has OpenMP pragmas, allowing a fair comparison with our OpenMP back-end.
For the HiSPEET kernels, we averaged the latency of 50 runs on 8192 elements each, which exhibit per-element parallelism and vectorization opportunities.

\subsection{Sequential Performance}\label{sec:single-thread-perf}

Concerning the \texttt{taumol\_sw} kernel, our naive CPU lowering scheme (cf. \cref{fig:cpu-minimap}) was able to outperform the \gls{wrf} baseline by a slight margin, using just our idiomatic code.
For reference, we also tested our idiomatic \NumPy{} kernel running the same workload, which took about \(\ms{12}\).

For a more representative test, our collaborators also ran a full \gls{wrf} prediction run on their HPC system using our \gls{theLang} plugin kernel.
While detailed measurements as provided in \cref{tab:results} were not feasible, an impact of below \(0.5\,\%\) on the total runtime was observed, including plugin overhead.
Note that the \texttt{taumol\_sw} kernel accounts for approximately \(4\,\%\) of the total prediction runtime.
Our collaborators also performed a statistical evaluation of the forecast result, which yielded a negligible deviation.
Although the kernel latency depends on the atmospheric state, and thus input data, this result supports the claim that \gls{theLang} matches the performance of the hand-written Fortran implementation.

The \texttt{inv\_helm} and \texttt{elliptic\_r} kernels both benefit from the reduction lifting described in \cref{sec:ekl-middle-end}.
This lifting is the main optimization that the expert-optimized code contains, which our \gls{dsl} compiler easily automates.
Due to aggressive vectorization and buffer sharing optimizations, the \texttt{inv\_helm} kernel's performance was further improved by \theCompiler{}.

The remaining three kernels contain a variety of reductions over subscripted stack expressions, which \theCompiler{} passes through to \gls{fpga} targets.
In their normalized form (cf. \cref{sec:ekl-middle-end}), these expressions are amenable to hardware synthesis via muxers and pipelining.
However, with our naive CPU lowering, this leads to unnecessary computations with bad locality.
To achieve acceptable single-thread performance, additional target specialization is necessary.

\subsection{Parallel Performance}

Our OpenMP back-end uses existing \gls{mlir} infrastructure to emit work sharing and reduction loops.
This requires no modifications to the \gls{dsl} code by the user, but relies heavily on the optimization capabilities of the middle-end dialects we reuse.
The \texttt{flang} Fortran compiler uses the same \gls{mlir} OpenMP facilities, but not the other dialects we use.
Expecting to be limited by the single-thread performance of \theCompiler{}, we still see a substantial speed-down for all our kernels, compared to the OpenMP baseline.

Bufferization, i.e., the assignment of memory buffers to tensor values, was found to be the main cause of this.
Bufferization is implemented by a piece of \gls{mlir} infrastructure that we reuse.
In the single-threaded case, currently existing analyses and optimizations are able to, e.g., eliminate spurious copies and allocations.
When lowering to OpenMP, the parallel nature of loops requires a different bufferization scheme, which is not established by existing passes.
In addition, the dynamic allocations de-facto serialize some parallel regions.
To generate efficient parallel implementations, \theCompiler{} requires an improved bufferization mechanism.

Other deficiencies include optimization hazards that are created around outlined OpenMP regions.
CPU-specific lowerings, as mentioned above, also become more relevant in the parallel case.
Specifically, we identified the need for immutable references to intermediates resulting from certain subscripts that can be (re-)used in parallel.

\subsection{\glsentryshort{fpga} Offloading}

While \gls{wrf}'s SPMD use of the \gls{ckd} kernel is not particularly amenable to offloading due to \gls{wrf}'s software architecture, we evaluated this for an \gls{fpga} target.
We used \gls{hls} to implement our operators, using Bambu~\cite{2021_Bambu} and CIRCT\footnote{\url{https://circt.llvm.org/}}, as both allow us to use \gls{mlir} as an input.
The \texttt{dfg-mlir} project allows us to model our kernels as \glspl{kpn} that replace memory buffers with streaming channels.
Etna~\cite{2024_Etna} is able to consume this representation and generate a memory-attached accelerator, complete with host driver code, our plugin can link against.
This flow requires the process nodes to be synthesized with AXI stream interfaces first.

While Bambu can synthesize the individual dataflow nodes, it can not implement dataflow like other \gls{hls} tools.
Since on-device dataflow is not yet supported by Etna, we could not fully automate the Bambu flow.
We also found that CIRCT was not mature enough to optimize the finite state machines created by the hardware lowering of \texttt{dfg-mlir}, leading to poor pipelining compared to Bambu.

Resorting to manual translation of \texttt{dfg} dialect features to dataflow pragmas for Vitis HLS, we were able to produce an accelerator for \texttt{taumol\_sw} without \gls{dsl} code changes.
We achieved a design latency of \(\approx 45\,\mu{}s\), which we believe is achieved in part due to burst memory accesses, enabled by the layout manipulations in the \theDialect{} dialect.
As a baseline, we synthesized AER's modernized \gls{ckd} kernel for GPUs, and achieved a design latency of \(\approx 600\,\mu{}s\).

\section{Conclusion}

While the promise of \gls{mlir} lies in reusability, practical \gls{dsl} implementations still require a lot of development effort and careful design to achieve it.
We demonstrated that feature-rich \gls{dsl} compilers can be implemented in \gls{mlir} without resorting to bespoke or legacy implementations.
In the process, we showed that specific low-friction \gls{mlir} extensions, such as lexical locations and type checking, greatly simplify front-end creation and promote reuse.
We created an open-source \gls{mlir} patch that adds subtyping and dialect-agnostic type checking as a first step towards \gls{mlir}-native front-ends.

By applying it all to a rather complex sample application, we showed that the promised advantages of \gls{mlir} are attainable even in its still limited ecosystem.
In fact, this reuse of established techniques and abstractions was uniquely made possible by \gls{mlir} alone.
However, we also faced some limitations that the current machine learning focused \gls{mlir} ecosystem poses to other uses.
We intend to use \theCompiler{} to identify and address more such issues to improve \gls{mlir}.

\begin{acks}
This work is partially funded by the EU Horizon 2020 Programme under grant agreement No 957269 (EVEREST) and by the EU Horizon Europe Programme under grant agreement No 101135183 (MYRTUS). Views and opinions expressed are however those of the author(s) only and do not necessarily reflect those of the European Union. Neither the European Union nor the granting authority can be held responsible for them.
We thank our colleagues at the CIMA Research Foundation for providing us with the test configuration and dataset, and helping with and performing our \gls{wrf} evaluation.
\end{acks}

\bibliographystyle{ACM-Reference-Format}
\bibliography{main}

\end{document}